\documentclass[sigconf,nonacm]{acmart}

\AtBeginDocument{%
  }

\usepackage[most]{tcolorbox}
\usepackage{graphicx}
\usepackage{subcaption}
\usepackage{float}
\usepackage{enumitem}
\tcbset{
  colback=gray!5!white, 
  colframe=gray!50!black, 
  colbacktitle=black,
  coltitle=white,
  fonttitle=\bfseries\footnotesize,
  boxrule=0.5pt, 
  arc=3pt, 
  left=6pt, 
  right=6pt, 
  top=2pt, 
  bottom=2pt, 
  boxsep=4pt,
  enhanced,
}
\usepackage{cleveref}

\setcopyright{acmlicensed}
\copyrightyear{2025}
\acmYear{2025}
\acmDOI{XXXXXXX.XXXXXXX}
\acmConference[ICAIF `25]{ACM AI in Finance}{Nov 15--18,
  2025}{Singapore}
\acmISBN{978-1-4503-XXXX-X/2018/06}

\begin{document}


\title{Explainable Federated Learning for U.S. State-Level Financial Distress Modeling}

\author{Lorenzo Carta}
\email{cartal@rpi.edu}
\orcid{0009-0005-5610-9093}
\affiliation{%
  \institution{Rensselaer Polytechnic Institute}
  \city{Troy}
  \state{New York}
  \country{USA}
}

\author{Fernando Spadea}
\email{spadef@rpi.edu}
\orcid{0009-0006-4278-3666}
\affiliation{%
  \institution{Rensselaer Polytechnic Institute}
  \city{Troy}
  \state{New York}
  \country{USA}
}

\author{Oshani Seneviratne}
\email{senevo@rpi.edu}
\orcid{0000-0001-8518-917X}
\affiliation{%
  \institution{Rensselaer Polytechnic Institute}
  \city{Troy}
  \state{New York}
  \country{USA}
}









\begin{abstract}
We present the first application of federated learning (FL) to the U.S. National Financial Capability Study, introducing an interpretable framework for predicting consumer financial distress across all 50 states and the District of Columbia without centralizing sensitive data. Our cross-silo FL setup treats each state as a distinct data silo, simulating real-world governance in nationwide financial systems. Unlike prior work, our approach integrates two complementary explainable AI techniques to identify both global (nationwide) and local (state-specific) predictors of financial hardship, such as contact from debt collection agencies. We develop a machine learning model specifically suited for highly categorical, imbalanced survey data. This work delivers a scalable, regulation-compliant blueprint for early warning systems in finance, demonstrating how FL can power socially responsible AI applications in consumer credit risk and financial inclusion.

Our open source code repository is available at: \url{https://github.com/brains-group/xfl-for-loan-eligibility}.
\end{abstract}




\maketitle

\section{Introduction}

\begin{figure}[t]
  \centering
  \includegraphics[width=\linewidth]{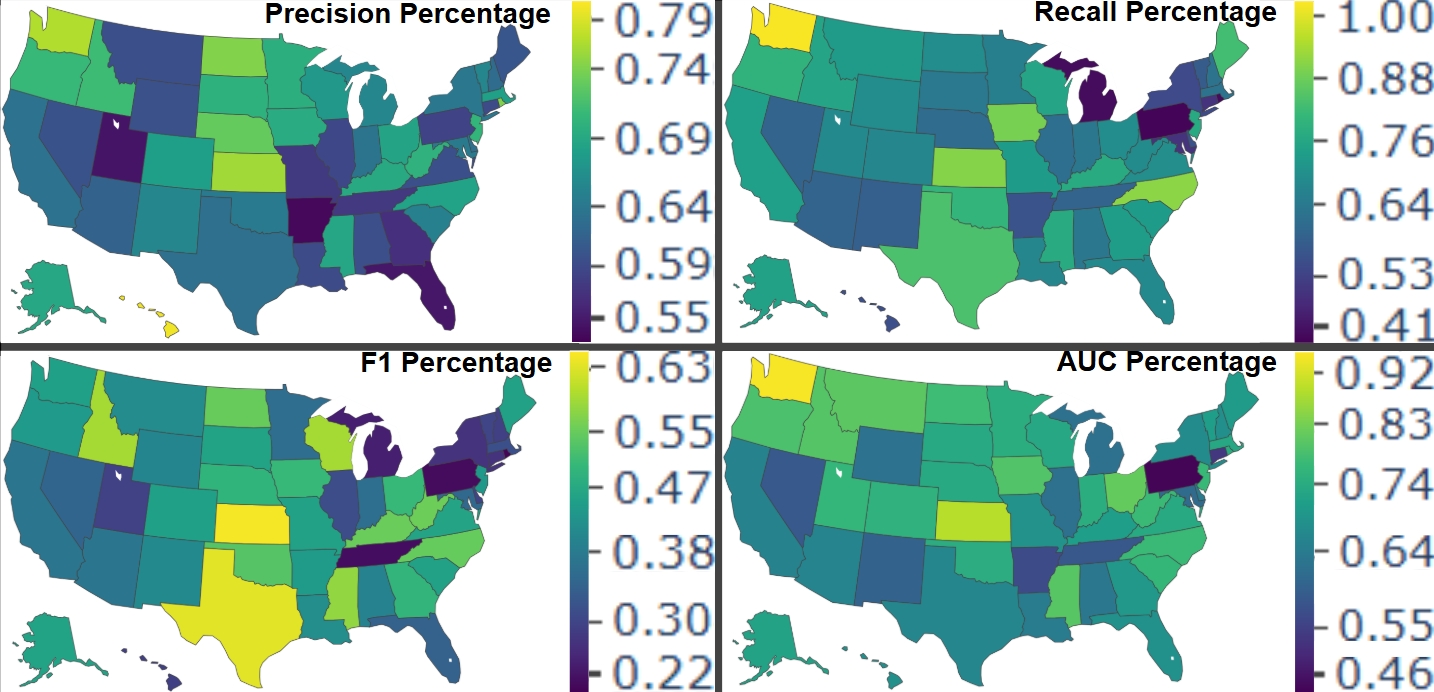}
  \caption{State-Level Evaluation of Federated Model Performance. \normalfont\small  {Choropleth maps showing the performance of the global federated model across U.S. states using four evaluation metrics: precision, recall, F1-score, and AUC. Each state acts as an independent client in the cross-silo FL setup, revealing heterogeneity in model effectiveness and highlighting regional variation in financial distress prediction.}}
  \label{fig:choropleth}
\end{figure}

Financial distress, often culminating in contact from a debt collection agency (DCA), is a leading indicator of economic instability and a precursor to long-term credit damage, bankruptcy, and social hardship. In the U.S. alone, tens of millions of consumers experience this annually~\cite{urban2022,FonsecaStrairZafar2017}. For financial institutions, regulators, and social welfare organizations, early identification of individuals at risk is critical to enabling targeted, proactive interventions that mitigate harm and promote financial inclusion.
While predictive modeling has shown great promise for anticipating such outcomes, real-world deployment in financial services remains constrained by two key factors: (1) the sensitive nature of personal financial data makes centralizing consumer-level information a regulatory and ethical challenge, and (2) black-box machine learning models undermine trust, particularly in high-stakes financial decision-making. Bridging these gaps requires techniques that safeguard data sovereignty while supporting interpretability.

In this work, we present a novel application of federated learning (FL) to predict a key signal of consumer financial distress, whether a person has been contacted by a DCA, based on demographic and behavioral data from the U.S. National Financial Capability Study (NFCS)~\cite{valdes2024finra}, a large-scale, representative survey managed by the FINRA Foundation. Rather than pooling data centrally, we simulate cross-silo FL across U.S. states, enabling localized state-level model development while capturing regional variation in financial hardship dynamics.
To support actionable insights, we augment our approach with explainable AI (XAI) techniques, integrating SHAP~\cite{sundararajan2020many} and Owen~\cite{kotsiopoulos2023approximation} values for richer explanations, identifying both nationwide and state-specific predictors of distress. 
Our findings show that our combined technique using FL and XAI surfaces interpretable patterns that can inform risk assessments, financial counseling strategies, and policy design.
To our knowledge, this is the first such federated approach applied to financial distress prediction on survey data across U.S. states, combining highly interpretable models with cross-state granularity.

Although our study focuses on U.S. survey data, the framework is broadly applicable to global financial ecosystems, particularly in countries with strong privacy regulations (e.g., GDPR in the EU) or fragmented data institutional infrastructures. In regions where credit bureaus are nascent or incomplete, government surveys or mobile financial data could serve as analogs to power similar decentralized risk prediction tools. FL enables such models to be trained across institutions (e.g., microfinance lenders, public sector agencies) or geographies without violating data sovereignty or exposing individuals to surveillance.

\subsection{Contributions}

The primary contributions of this work are:

\begin{enumerate}
    \item \textbf{Design and evaluation of a novel FL framework for financial distress prediction.} We designed and evaluated a cross-silo FL pipeline to predict DCA risk using data from the NFCS~\cite{valdes2024finra}. The model was trained on data partitioned across the 50 U.S. states and the District of Columbia.

    \item \textbf{Development of a tailored architecture for handling imbalanced survey data in FL settings:} We developed a solution for the challenges of real-world survey data by combining an 8-layer Highway Network~\cite{greff2016highway} with a targeted class weighting strategy.

    \item \textbf{Application of multi-faceted XAI for comprehensive model explanation:} We applied and compared SHAP~\cite{sundararajan2020many} and Owen value~\cite{kotsiopoulos2023approximation} techniques to interpret model behavior at both global (nationwide) and local (state-specific) levels. This analysis identifies critical risk factors like age and income and demonstrates how distributing the total model prediction value among features in a way that respects their cooperative contributions offers a more intuitive explanation for interdependent, categorically binned features.


\end{enumerate}

\section{Related Work}

\paragraph{\textbf{Financial Distress Prediction and Credit Risk Modeling}}

Predicting financial distress, whether personal default or corporate bankruptcy, has been a long-standing focus in both finance and machine learning~\cite{altman2017financial,keasey2019financial}. Classical models such as logistic regression have dominated this space for decades due to their interpretability and alignment with financial theory, consistently identifying key risk factors like liquidity, solvency, and firm size~\cite{yang2025financial}. 
In the consumer domain, datasets like the FINRA Foundation’s NFCS have been used to study individual-level financial distress at a macro scale~\cite{valdes2024finra}.
However, the the sensitivity and governance restrictions of the underlying financial data often limit its use to centralized or pre-aggregated analyses, preventing more granular, jurisdiction-aware modeling. This is especially limiting given that access to even more sensitive or fine-grained data, such as transaction histories, credit scores, or behavioral signals, at the local level could yield deeper insights into region-specific patterns of financial distress. These signals are often obscured or lost in centralized datasets due to anonymization, aggregation, and privacy-preserving constraints.
Our work addresses this gap by applying FL to support fine-grained, cross-state modeling, while aligning with the data sovereignty and access control policies typical of large-scale aggregated datasets. Importantly, we demonstrate that even when using publicly available survey data like FINRA’s, federated training yields good predictive performance, highlighting the potential of this approach for applications involving richer and more sensitive data sources.

\paragraph{\textbf{Decentralized and Federated Learning in Finance}}

First introduced by \citet{mcmahan2017communication}, FL is a decentralized machine learning paradigm that enables multiple parties to collaboratively train a model without sharing raw data.
In the financial services domain, FL is seen as a promising solution to the “data silo” problem where sensitive customer data cannot be centrally pooled, for applications such as financial crime detection~\cite{khan2024fed,aljunaid2025secure} and credit scoring~\cite{wang2024novel,jovanovic2024robust}.
\citeauthor{lee2023federated} (\citeyear{lee2023federated}) developed an FL prototype for credit risk modeling across smaller financial institutions, enabling them to “compete by training in a cooperative fashion” on a shared mortgage dataset~\cite{lee2023federated}.
Similarly, \citeauthor{wang2024novel} (\citeyear{wang2024novel}) proposed an FL method for credit scoring, addressing data imbalance concerns~\cite{wang2024novel}.
Our work builds on this foundation by applying cross-silo FL to a nation-wide social finance survey, treating each U.S. state as a silo, a scenario that, to our knowledge, has not been previously explored for financial distress prediction.

\paragraph{\textbf{Explainable AI for Financial Distress Models}}

Regulatory and practical constraints in finance require that predictive models offer clear explanations, especially in high-stakes decisions like credit approvals, where lenders must justify adverse actions. As a result, there is growing interest in integrating XAI into financial risk modeling. One strategy involves using inherently interpretable models, such as linear classifiers or decision rules, that provide transparent reasoning at the cost of reduced complexity~\cite{yang2025financial}.
Alternatively, post-hoc explanation techniques like Shapley Additive Explanations (SHAP)~\cite{sundararajan2020many} have been applied to complex models to retain both performance and interpretability. For example, \citeauthor{nguyen2024using} (\citeyear{nguyen2024using}) used SHAP to uncover the key financial drivers (e.g., leverage, cash flow) behind corporate distress predictions made by neural networks and gradient-boosted trees~\cite{nguyen2024using}. Similarly, \citet{aljunaid2025secure} propose an explainable FL model for bank fraud detection.
Additionally, \citet{jovanovic2024robust} integrate blockchain with explainable FL for credit risk modeling.
We extend this line of work by applying explainable FL to the domain of personal financial distress, a setting where interpretability is essential for both regulators and consumers. Our contribution is distinct in combining SHAP with Owen values, enabling interpretation of both global (national) and local (state-level) model behavior. This dual-layer explainability also addresses the challenge of making complex FL models interpretable for each data-holding client, particularly when categorical features exhibit strong interdependencies.

\section{Methodology}

\subsection{Dataset and Preprocessing}
\label{sec:dataset}

We used the FINRA Investor Education Foundation’s 2021 NFCS~\cite{valdes2024finra} for our analysis. This comprehensive dataset is derived from a 126-question multiple choice survey that captures various statistics such as income, education, and financial education from more than 25,000 U.S. respondents, with approximately 500 respondents per state, plus the District of Columbia. The raw data, available in comma-separated-value (CSV) format categorically numbers each respondent's multiple-choice answer (i.e., labeled as 0 for the first choice, 1 for the second, etc.). We extract 11 of the 126 features, along with 1 designated learning objective. To facilitate binary classification, respondents who gave a non-binary response to the chosen learning objective were filtered out. 

\paragraph{\textbf{Prediction Goal}}
The specific learning objective extracted came from question G38 of the NFCS survey: "Have you been contacted by a debt collection agency in the past 12 months?" Respondents answered with a `Yes', `No', `Don’t know', and `Prefer not to say'; the latter 2 of which were filtered out for binary classification. A notable challenge with this prediction goal was the significant class imbalance: having over 78.5\% of respondents answering `No' and only 18\% answering `Yes'. This imbalance initially led to the model collapsing into always predicting `No', resulting in high accuracy at the cost of trivial predictions; thus, the implementation of class weighting was crucial to address this issue and enable effective learning of the minority class.

\paragraph{\textbf{Feature Selection}}
Of the 126 available features in the NFCS dataset, only 12 were selected for this model. This decision was driven by concerns regarding the large number of subjective questions and the potential for biased data within the full survey. Out of all the features, the 12 selected ones were shown to be the most objective and relevant features, focusing on less sensitive financial and demographic information such as income range and living arrangements. Additionally, they represent information that could be realistically and ethically queried from willing participants in alternative data collection scenarios.

\paragraph{\textbf{Data Cleaning and Encoding}}
All features used in the model were categorical and derived from a questionnaire. During data loading, certain responses were flagged for exclusion (e.g., non-binary answers to a binary question) and removed from the model's learning structure.

\subsection{FL Setup}
\label{sec:federated}
Our model employs a cross-silo, Flower-FL setup~\cite{beutel2020flower}. The dataset is initially partitioned into a training and testing set, each holding 80\% and 20\% of the raw data respectively. The training data is further divided among 51 separate clients, each representing one of the 50 U.S. states and the District of Columbia. These clients independently train their local models and send their updates to a global server, which then utilizes Federated Averaging (FedAvg)~\cite{mcmahan2017communication} to aggregate the local updates and compute a single, server-wide model update. After the training is complete, the final global model's performance is evaluated against the partitioned testing data, and overall metrics are extracted. Furthermore, the extracted metrics are compared to a centralized model with identical attributes to demonstrate that FL achieves similar performance. Additionally, to show that the state-specific results with FL are better than without, we train local models as a baseline to compare against. The FL state-wide data partitioning is made possible due to the inclusion of the `STATEQ' variable within the NFCS dataset, which numerically represents each respondent's state of origin alphabetically (e.g., 1 = Alabama, 2 = Alaska, etc).

\subsection{Model Architecture}
\label{sec:model-architecture}
We employ an 8-layer highway network because its gated architecture enables training of deep neural networks by mitigating vanishing-gradient problems~\cite{greff2016highway}.
Our model design, inspired by highway networks’ success in other domains (e.g. sequence labeling~\cite{liu2018empower} and language modeling~\cite{kurata2017language}), is well-suited to our tabular survey data with many categorical features. The gating mechanism allows the model to effectively learn complex feature interactions while skipping unnecessary transformations, which we found beneficial and yielded stronger results than standard architectures (demonstrating an approximate 9\% increase in accuracy to a standard LeakyReLu model), given the high dimensionality and sparsity of the NFCS features.

Furthermore, to address the substantial class imbalance in the learning objective, where positive cases (i.e., individuals contacted by a DCA) are significantly underrepresented, class weighting was applied to prevent the model from biasing toward the majority class. The weighting factor was determined empirically after training the model multiple times using various different factors. After evaluating several configurations, we found that increasing the standard minority class weight by a factor of 1.1835 yielded the best balance between precision and recall, enabling the model to learn meaningful patterns from the minority class while maintaining stable overall performance.
Our model implementation also incorporates various training tools, including the Adam optimizer~\cite{kingma2014adam}, and early stopping.

\vspace{-1em}
\subsection{Evaluation Metrics}
\label{sec:eval}
To assess the overall training results, this model utilizes four distinct evaluation metrics:
\begin{itemize}[leftmargin=*, noitemsep, topsep=0pt, parsep=0pt, partopsep=0pt]
    \item Precision, which denotes the accuracy of the model's positive predictions across the entire testing dataset.
    \item Recall, which represents the model's accuracy in identifying all relevant instances of the positive minority class (i.e., individuals who were contacted by a DCA).
    \item F1-Score, which provides the harmonic mean of precision and recall and offers a balanced measure of the model's accuracy.
    \item Area Under the Receiver Operating Characteristic Curve (AUC), which quantifies the model's ability to distinguish between the two classes. 
\end{itemize}

In addition to these metrics, we also analyze the individual feature summaries used to discern feature importance, using SHapley Additive exPlanations (SHAP)~\cite{sundararajan2020many} and Owen values~\cite{kotsiopoulos2023approximation}. These two methods, while related, employ distinct attribution mechanisms: SHAP based on marginal contribution across permutations, and Owen values based on structured coalition games, which can lead to both overlapping and divergent insights about feature importance.

\begin{figure*}[!ht]
    \centering

    \begin{subfigure}{0.7\textwidth}
        \centering
        \includegraphics[width=0.95\linewidth]{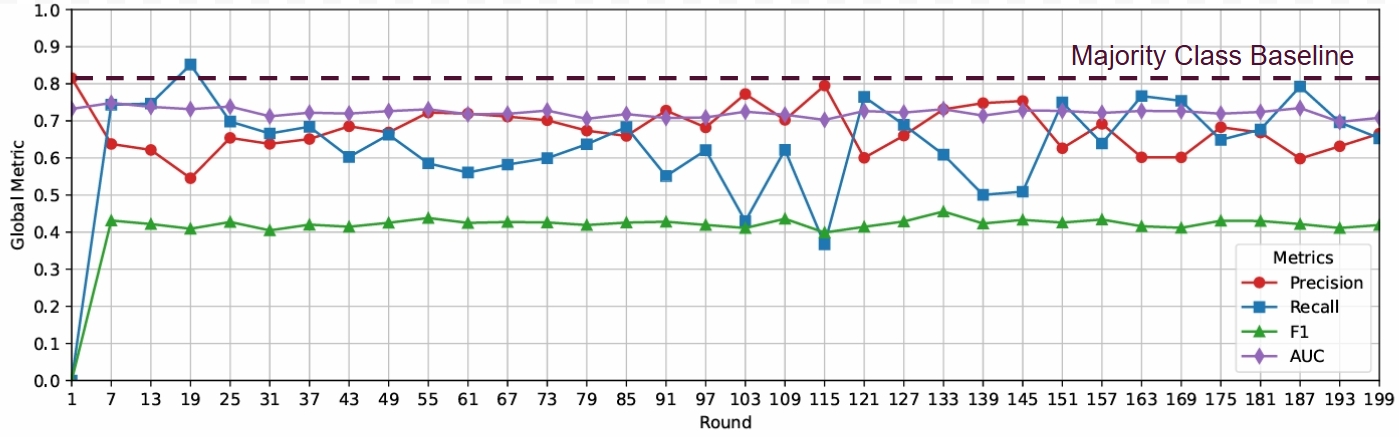}
        \caption{Evolution of key evaluation metrics over 200 training rounds.}
        \label{fig:metrics}
    \end{subfigure}
    \begin{subfigure}{0.3\textwidth}
        \centering
        \includegraphics[width=0.96\linewidth]{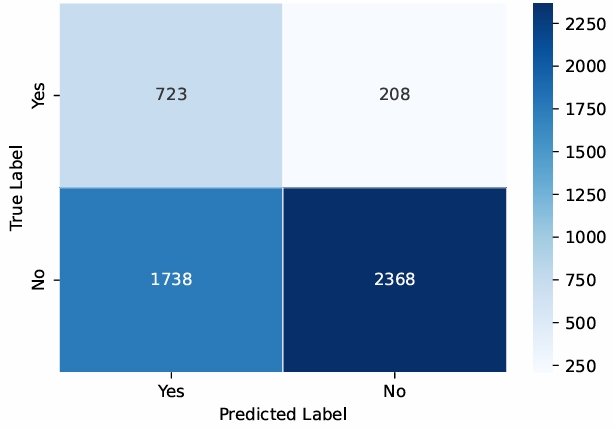}
        \caption{Confusion matrix}
        \label{fig:confusion-matrix}
    \end{subfigure}

    \caption{Global performance metrics of the federated model over training rounds and final confusion matrix on the test set.}
    \label{fig:combined-metrics}
\end{figure*}

\vspace{-1em}

\section{Experimental Results}


\subsection{Global Model Performance}

 
As can be seen in \Cref{fig:metrics}, the global model converges rapidly in the early rounds, with relatively stable performance thereafter, with only minor trade-offs between precision and recall, 
However, throughout the training, a seemingly negative correlation can be observed between precision and recall, which have an inversely proportional relationship. This trade-off arises because similar feature profiles may map to both positive and negative outcomes, so optimizing for one metric often comes at the cost of the other.

\Cref{fig:confusion-matrix} further reflects the challenge of class imbalance in the global model. Despite this, the model achieved 61.4\% precision while maintaining 77.7\% recall, indicating that it successfully identified most true positive cases while avoiding excessive false positives.

As mentioned in \Cref{sec:dataset}, the dataset was heavily imbalanced, with over 78.5\% of samples belonging to the negative class.

As a consequence of the empirically-tuned class weighting, the model achieved a more balanced performance between overall precision and recall, with some training rounds demonstrating over 65\% accuracy in both metrics. 

\subsection{Local (State-Level) Model Performance}

\begin{figure}[!htbp]
    \centering 

    \begin{subfigure}{0.48\textwidth}
        \centering
        \includegraphics[width=\linewidth]{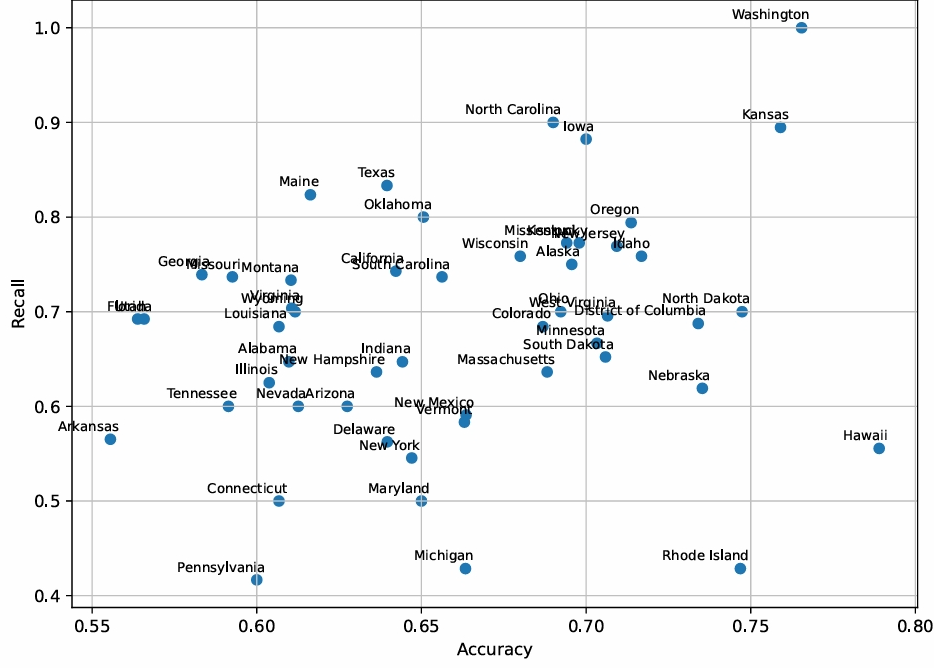}
        \caption{State-Level Precision vs. Recall}
        \label{fig:ac-re-scatter}
    \end{subfigure}
    \hfill 
    \begin{subfigure}{0.48\textwidth}
        \centering
        \includegraphics[width=\columnwidth]{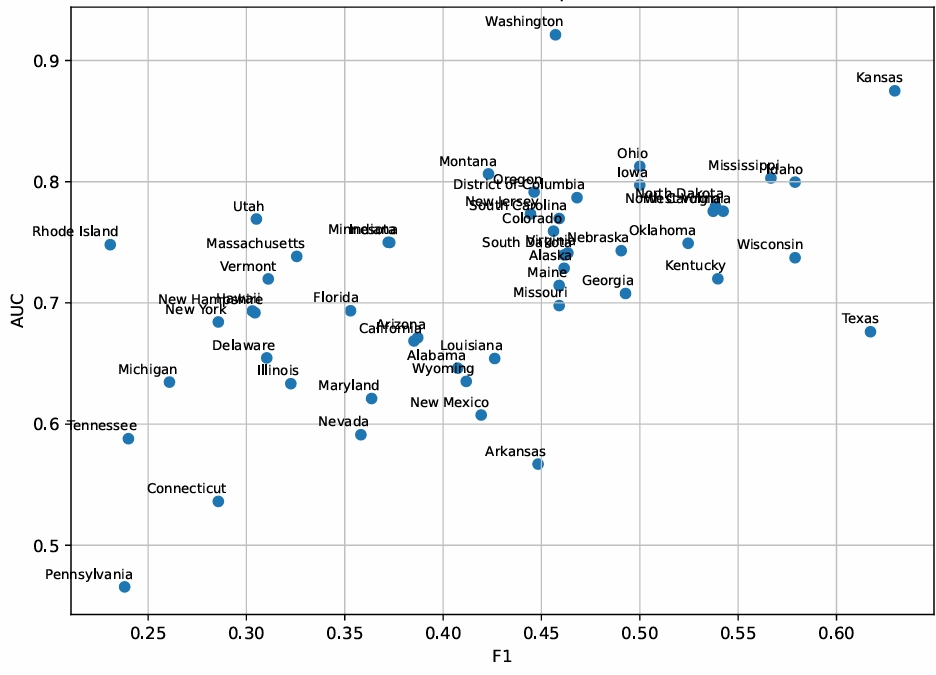}
        \caption{State-Level F1 Score vs. AUC}
        \label{fig:f1-au-scatter}
    \end{subfigure}

    \caption{State-level scatterplots of model performance metrics across all 51 clients (50 states + District of Columbia).
    }
    \label{fig:combined-scatter}
\end{figure}


While the global federated model is trained collaboratively across all clients, its performance can vary significantly when evaluated independently on each state’s local test data. Below, we analyze how well the shared global model generalizes to individual states.

Each point in \Cref{fig:ac-re-scatter} represents a state's performance, highlighting variation in the model’s ability to identify true positives (recall) versus its selectivity (precision). Notable outliers include Washington (high recall and precision) and Hawaii (high precision, low recall), reflecting diverse local patterns.

Contrary to initial expectations, the data in \Cref{fig:ac-re-scatter} did not appear inversely proportional, but instead appears to be fairly clustered with a slight positive correlation in its distribution. 
Most states appear to reside within this cluster, excluding some outliers such as Washington (which consistently outperformed all other clients’ recall), Hawaii, Rhode Island, Michigan, and Pennsylvania. In general, every client produced a favorable precision > 55\%, and a decent recall accuracy: with only 3 states demonstrating suboptimal results. Overall, it is very interesting to observe such variation within the scatterplot, where some states can have very similar precision but vastly different recalls (e.g., Washington compared to Hawaii), illustrating that the two metrics do not necessarily correlate to each other.


Each point in \Cref{fig:f1-au-scatter} indicates how well the model balances precision and recall (F1) and its overall discriminative ability (AUC). While most states cluster in a moderate performance range, states like Washington and Kansas show strong performance on both metrics.
\Cref{fig:f1-au-scatter} followed our initial expectations and demonstrates a strong proportional relationship between the two metrics. 
Overall, all clients except for Pennsylvania demonstrate an AUC above 50\%, but an F1 score below 65\%. 
The low F1 score highlights the inherent difficulty in precisely identifying all instances of DCA contact within an imbalanced real-world dataset, even with applied class weighting.

Furthermore, per-client metric results for the final round can also be observed in \Cref{fig:choropleth}, which shows the same results as in \Cref{fig:ac-re-scatter} and \Cref{fig:f1-au-scatter}, but in a more geographically evident manner, with states like Washington and New Jersey colored to represent a high precision and states like Texas and Mississippi showing a higher F1 score.

\subsection{Centralized and Local Model Comparison}

\begin{figure*}[!ht]
    \centering
    \begin{subfigure}{0.65\textwidth}
        \centering
        \includegraphics[width=0.95\linewidth]{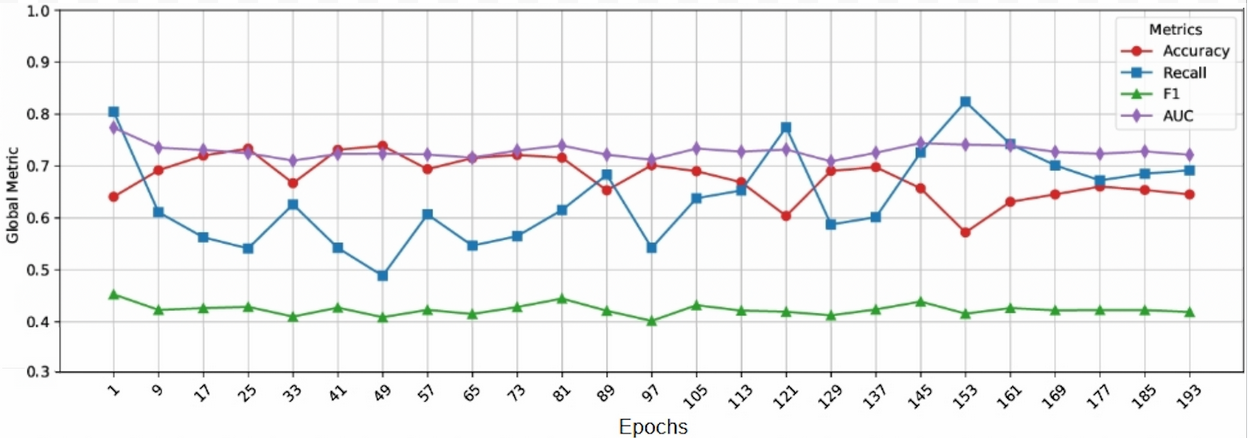}
        \caption{Centralized model: evaluation metrics over 200 training epochs.}
        \label{fig:metricsC}
    \end{subfigure}
    \begin{subfigure}{0.32\textwidth}
        \centering
        \includegraphics[width=0.96\linewidth]{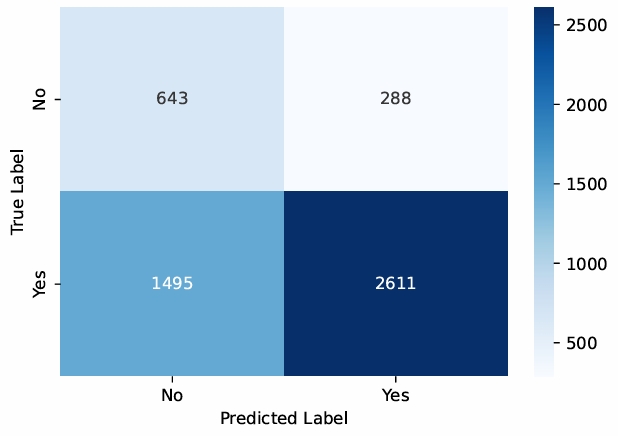}
        \caption{Centralized model: Confusion Matrix.}
        \label{fig:confusion-matrixC}
    \end{subfigure}
    \caption{Centralized Model Performance}
    \label{fig:centralized-model} 
\end{figure*}

To establish a benchmark for FL performance, we additionally trained a centralized version of the DCA prediction model. We incorporated the exact same parameters as those that were used in the FL setting, with the only change being the empirical adjustment of the minority class weight factor from 1.1835 to 1.182 in order to account for the un-intuitive differences in data modeling. As can be seen in \Cref{fig:metricsC}, over the course of 200 epochs, the centralized model exhibited convergence behavior very closely aligned with that of the federated model, with its performance trends nearly identical across all 4 metrics. After computing the confusion matrix after the final epoch (\Cref{fig:confusion-matrixC}), we found the centralized model achieved an F1 score of 42.4\% and an AUC score of 74.1\%. Compared to the federated model's scores of 42.2\% and 71.4\% in F1 and AUC respectively.

Furthermore, we trained a localized model across each individual state and took the average metric values for comparison. An important note to address is the fact that some states had very biased datasets, with a couple of states (NJ, ND) not even having any positive data points, which majorly impacted the average metric results. Overall, over 20 epochs of training, the average per-state centralized training model reached an average F1 score of 33.31\%, and an AUC score of 69.96\%, highlighting a much less favorable overall performance compared to FL, which was trained to be generalized to every state. These findings highlight that FL not only matches the accuracy benchmarks of centralized learning in single-model cases, but also surpasses it in aggregate performance while preserving data privacy and security.

\subsection{Communication Cost Analysis}

While our cross-silo FL framework preserves data locality, it necessarily incurs communication overhead due to repeated synchronization between the central server and participating clients. To manage this cost while ensuring representative learning, we adopt a partial participation strategy: in each communication round, 12 out of the 51 clients (approximately 25\%) are randomly selected for training. This participation rate follows common practice in the FL literature~\cite{mcmahan2017communication}, balancing the diversity of updates with reduced per-round bandwidth requirements.
Since clients do not use the model for local tasks when they are not actively training, the server only needs to send the global model to the 12 selected clients and receive their locally updated models.
Therefore, in each round, the server sends the global model to 12 clients and receives their locally updated models. Given that our 8-layer highway network occupies approximately 1,078 KB, each round requires 12,936 KB in upload and an equal amount in download, yielding a total of 25,873 KB per round. Over 200 training rounds, the total communication volume amounts to roughly 4.94 GB.

In comparison, a naïve approach in which all 51 clients receive and return models in every round would result in a communication cost of approximately 12.95 GB. Thus, our strategy achieves more than a 60\% reduction in bandwidth consumption without compromising model performance or fairness across clients.

\subsection{Feature Importance}

To understand the contribution of each input feature to the model's predictions, we analyzed feature importance using both SHapley Additive exPlanations (SHAP)~\cite{sundararajan2020many} and Owen values~\cite{kotsiopoulos2023approximation}, as introduced in \Cref{sec:eval}. Our analysis revealed varying degrees of influence among the 11 selected features.
The \texttt{Gender/Age} bin feature was included twice, once grouped by gender and once by age, to separately assess the contribution of each component within the binned categorical structure and highlight how feature interactions influence model attribution across explanation methods 
in classifying whether a respondent was contacted by a DCA.

\begin{figure}[!htbp]
    \centering

    \begin{subfigure}{\columnwidth}
        \centering
        \includegraphics[width=\linewidth]{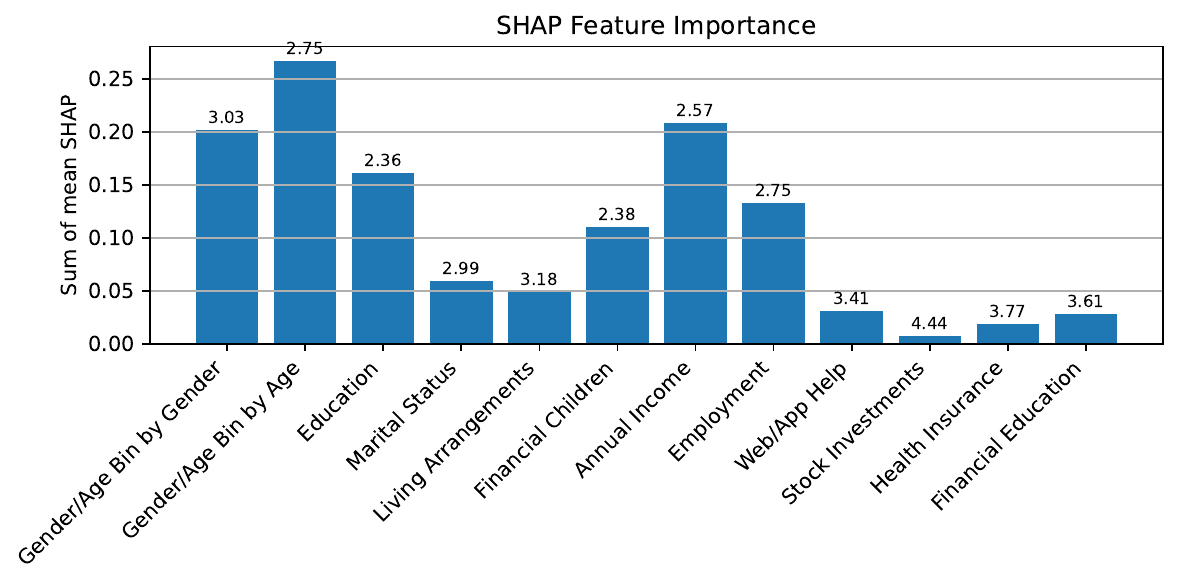}
        \caption{SHAP value summary showing the average contribution of each input feature to the model’s predictions.}
        \label{fig:shap-feature}
    \end{subfigure}
    
    \vspace{1em} 

    \begin{subfigure}{\columnwidth}
        \centering
        \includegraphics[width=\linewidth]{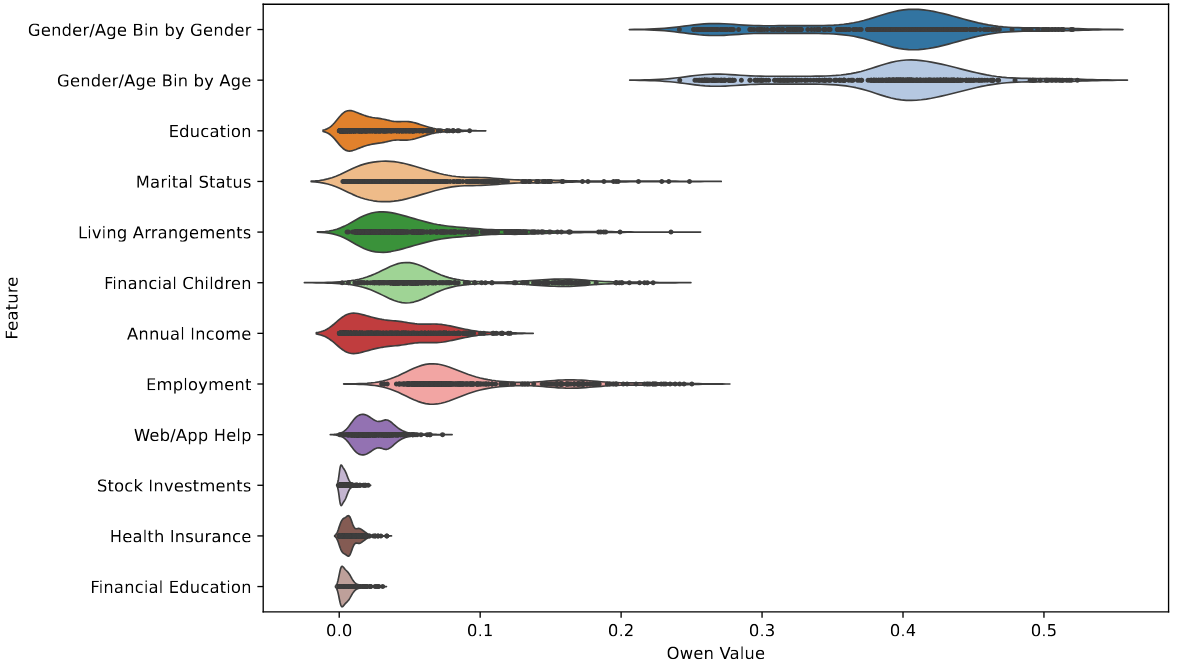}
        \caption{Owen value distribution plot, capturing the cooperative influence of each feature under interdependence.}
        \label{fig:owen-feature}
    \end{subfigure}

    \begin{subfigure}{\columnwidth}
      \centering
      \includegraphics[width=\linewidth]{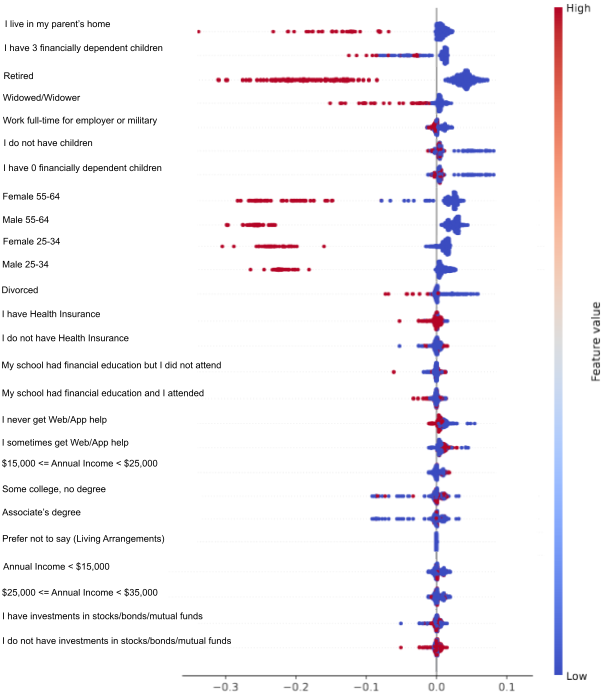}
      \caption{Grouped Owen value summary plot.}
      \label{fig:owen-grouped}
    \end{subfigure}
    
    \caption{Global feature importance analysis using SHAP and Owen values.}
    \label{fig:combined-feature-importance}
\end{figure}

\subsubsection{SHAP Analysis}

\sloppypar{
As can be seen in \Cref{fig:shap-feature}, demographic features such as \texttt{Gender/Age} bins and income exhibit the highest impact on the model's output, while features like \texttt{Financial Education} and \texttt{Stock Investments} contribute less.}
%
Following the \texttt{Gender/Age} bins, \texttt{Annual Income} was found to be a highly influential feature, with a mean SHAP value of approximately 0.16. This finding intuitively aligns with expectations, as income is directly related to financial capacity and the ability to manage debt. \texttt{Employment Status} (around 0.105) and \texttt{Education} (around 0.11) also demonstrated considerable importance, reflecting their well-established correlations with financial stability and access to resources.
Conversely, \texttt{Stock Investments} was identified as the feature with the lowest mean absolute SHAP value (approximately 0.02) among all analyzed features, indicating its minimal average contribution to the prediction. 

However, as demonstrated by the large standard deviation values and surprising difference between \texttt{Gender/Age Bin (Age)} and \texttt{Gender/Age Bin (Gender)}, despite only differing categorically, we conclude that SHAP values alone are not a very good metric to use when evaluating the largely categorical features in the NFCS dataset.

\subsubsection{Owen Value Analysis}

In contrast to SHAP values, Owen values offer a more consistent representation for categorical variables, showing stronger separation between key features (e.g., \texttt{Gender/Age}, \texttt{Employment}) and highlighting their relevance to model output, as can be seen in \Cref{fig:owen-feature}.
This is because Owen values, unlike SHAP, are designed to distribute the total model prediction value among features in a way that respects their cooperative contributions, making them robust when dealing with interdependencies often found in categorical data.


A crucial observation from the Owen value analysis (\Cref{fig:owen-feature}) is the equal contribution of \texttt{Gender/Age Bin (Gender)} and \texttt{Gender/Age Bin (Age)}, both showing Owen values concentrated around 0.5. This equality in Owen values provides a more intuitive and consistent representation of their combined importance, especially considering that they are categorical permutations derived from the same underlying demographic information. The discrepancy highlights how Owen values can be a better indicator for this type of binned categorical data, providing a more balanced view of interdependent feature groups.

Beyond the \texttt{Gender/Age} bins, the Owen values further clarify the relative importance of other features. Features such as \texttt{Employment Status}, \texttt{Marital Status}, and \texttt{Financial Children} demonstrated significant impact on the model, showing denser distributions of Owen values further from zero. These features intuitively carry substantial weight, as an individual's employment stability, marital status, and financial dependents can be thought to directly influence their household's financial obligations and capacity to manage debt.

In contrast, features like \texttt{Stock Investments}, \texttt{Financial Education}, and \texttt{Health Insurance} consistently exhibited Owen values clustered closer to zero, indicating a considerably lower impact on the model's predictions. This is also intuitively explainable, because although these aspects relate to an individual's broader financial landscape, it is not directly visible or immediately actionable information for a DCA.

\subsubsection{Bin-Level Owen Distributions}

To gain a granular understanding of how each feature influences the model's predictions, we analyzed the Owen values for specific hand-picked individual feature bins, as presented in \Cref{fig:owen-grouped}. These plots illustrate the distribution of Owen values across different categories within each feature.


\Cref{fig:owen-grouped} uses bee-swarm feature summary to plot the Owen summary values for each specific category or bin within a feature (e.g., ``Male 25-34'' from the \texttt{Gender/Age} feature, ``Retired'' from \texttt{Employment}, etc.), with the most interesting categories being selected for visualization. The spread and concentration of points along each line, colored by feature value (with red indicating positive respondents within the category and blue representing negative ones), demonstrate the degree to which each category influences the model's output. An important aspect to note is that Owen values only compare the data with itself, and thus does not produce a value when there is nothing to compare to. For example, as observed in the ‘Prefer not to say (Living Arrangements)’ row in \Cref{fig:owen-grouped}, categories that did not receive any positive (red) datapoints resulted in a zeroing of negative points, which could not be compared.

Consistent with the aggregate Owen value analysis in \Cref{fig:owen-feature}, the \texttt{Gender/Age} bins had the most pronounced influence on model predictions. Both male and female age categories displayed highly spread-out values across nearly all points in the bee-swarm plots in \Cref{fig:owen-grouped}. This suggests that these specific demographic profiles are strong indicators of whether an individual is likely to be contacted by a DCA. In addition to the \texttt{Gender/Age} bins, employment status also emerged as a key predictor: retired individuals exhibited significantly lower Owen values compared to non-retired individuals, while full-timers showed a less distinct but visually separable trend.

Marital status and health insurance status also had similar patterns of influence: certain marital status categories, such as divorced or widowed individuals, displayed similar bee-swarm distributions, with red points tending to show negative Owen values. Likewise, the presence of health insurance generally resulted, albeit with less effect, in a lower Owen score and vice versa.

\subsubsection{Features with Minimal Predictive Value}

In contrast, several features consistently had little to no impact on the model’s predictions, with their Owen values clustered around zero. These include:
\begin{itemize}[leftmargin=*, noitemsep, topsep=0pt, parsep=0pt, partopsep=0pt]
    \item Stock/Bond Investments: The wide spread of identically-colored points across both ends of the value spectrum suggests that this feature has minimal impact, which intuitively shows that DCAs generally lack access to individuals' investment data.

    \item Annual Income: Surprisingly, this feature did not strongly correlate with a person’s likelihood of being contacted by a DCA. This indicates that a person’s payment history and existing debt obligations are not closely tied to their earning potential.

    \item Financial Education: Despite some outliers, the general distribution of data points indicates that having formal financial education alone is not strongly associated with a reduced likelihood of being contacted by a DCA.
    In contrast, individuals who reported receiving financial help from dynamic sources such as websites or mobile apps exhibited greater separability in the model’s predictions, suggesting that fresh, accessible financial guidance at the point of need tends to be more impactful than static or outdated knowledge acquired through past educational experiences.
\end{itemize}

\begin{figure*}[!htbp]
    \centering

    \begin{subfigure}{0.32\textwidth}
        \centering
        \includegraphics[width=\linewidth]{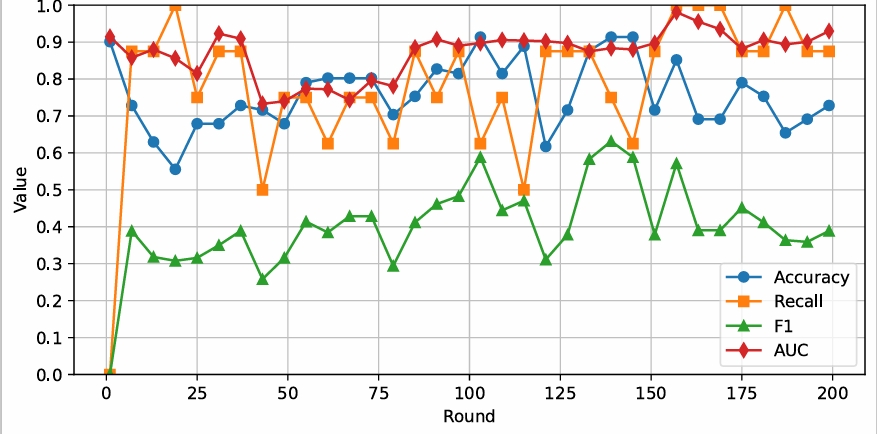}
        \caption{Per-Round Metrics of Washington}
        \label{fig:wash-metrics}
    \end{subfigure}
    \hfill 
    \begin{subfigure}{0.32\textwidth}
        \centering
        \includegraphics[width=\linewidth]{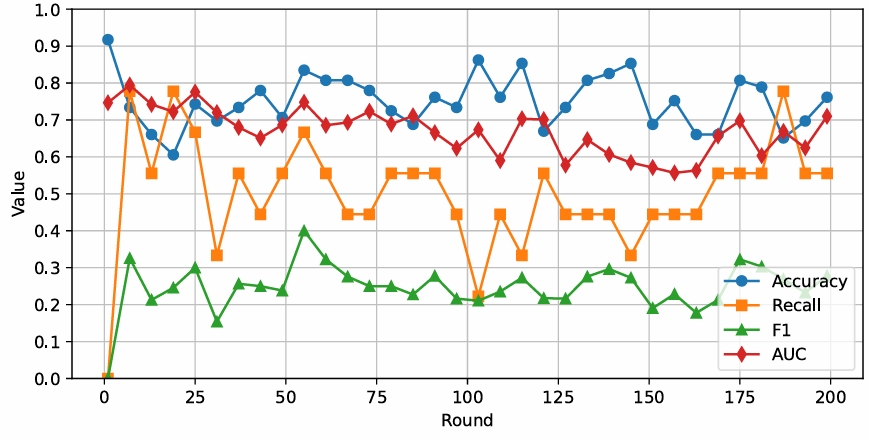}
        \caption{Per-Round Metrics of Hawaii}
        \label{fig:haw-metrics}
    \end{subfigure}
    \hfill 
    \begin{subfigure}{0.32\textwidth}
        \centering
        \includegraphics[width=\linewidth]{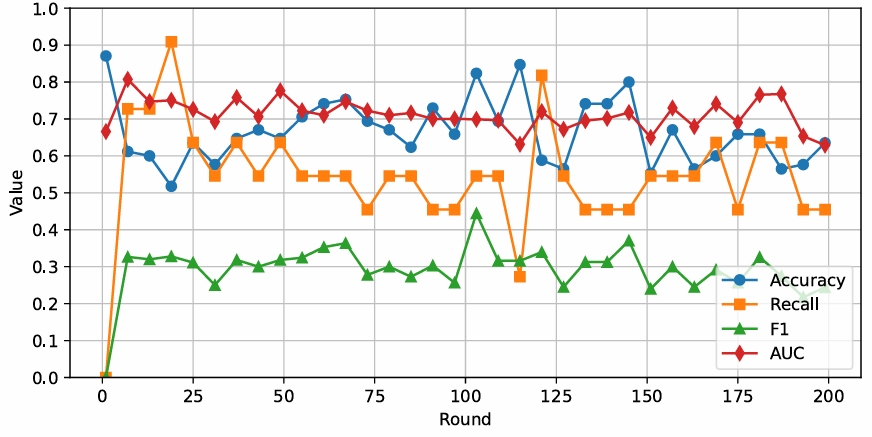}
        \caption{Per-Round Metrics of New York}
        \label{fig:ny-metrics}
    \end{subfigure}

    \vspace{1em} 

    \begin{subfigure}{0.32\textwidth}
        \centering
        \includegraphics[width=\linewidth]{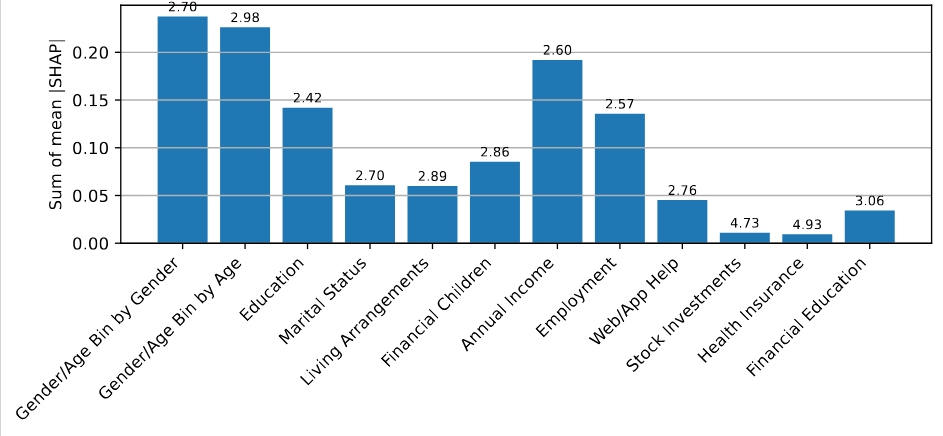}
        \caption{Washington SHAP feature importance}
        \label{fig:wash-feature}
    \end{subfigure}
    \hfill 
    \begin{subfigure}{0.32\textwidth}
        \centering
        \includegraphics[width=\linewidth]{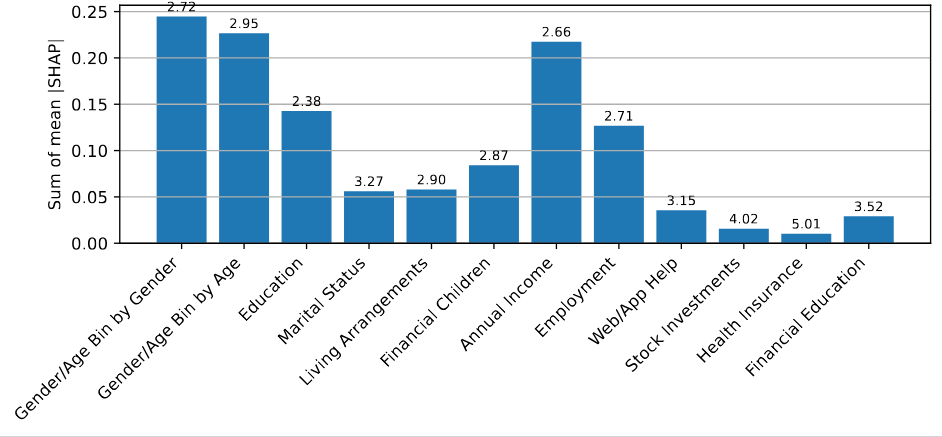}
        \caption{Hawaii SHAP feature importance}
        \label{fig:haw-feature}
    \end{subfigure}
    \hfill 
    \begin{subfigure}{0.32\textwidth}
        \centering
        \includegraphics[width=\linewidth]{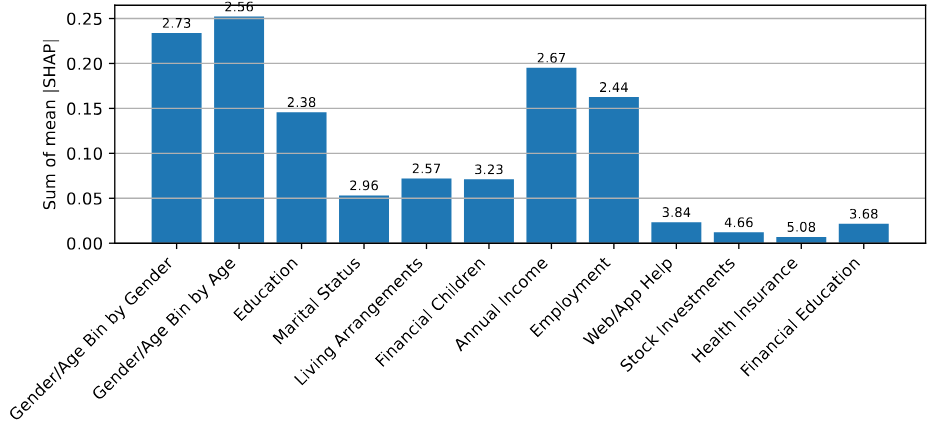}
        \caption{New York SHAP feature importance}
        \label{fig:ny-feature}
    \end{subfigure}

    \vspace{-2mm}
    \caption{A comparison of per-round training metrics and SHAP feature importance across Washington, Hawaii, and New York.}
    \label{fig:combined-state-analysis}
    \label{fig:state-comparison}
    \vspace{-3mm}
\end{figure*}

\subsection{Comparative Model Behavior Identifying Outlier and Representative States}

Throughout the training rounds the FL model was subjected to, a small subset of states consistently emerged as outliers in terms of their individual performance metrics. While the specific set of outlier states varied in every session (due to data heterogeneity and stochasticity of the training process), we chose to highlight two common outlier states, Washington and Hawaii, for which we calculated and plotted the individual state's performance metrics, as well as their SHAP feature summary.
We also generated the individual metrics and feature summary for New York, a populated state whose final metrics closely mirrored those of the global model. Finally, we compare each plot to provide a comprehensive view of the state-level model behavior as can be seen in \Cref{fig:state-comparison}. 


Looking at our first individual metrics in \Cref{fig:wash-metrics}, Washington consistently demonstrated exceptional performance, outperforming the global model in nearly every training round. Washington's local model consistently achieved high scores across all evaluation metrics, often reaching values close to 0.9 or higher during its training. This sustained high performance suggests that the data distribution within Washington were particularly well aligned with the learning objective.


As shown in \Cref{fig:wash-feature}, we observe a local feature importance profile that broadly mirrors the global model's trends (\Cref{fig:shap-feature}). Excluding some differences like the overperformance of \texttt{Gender/Age Bin (Gender)} and the underperformance of the \texttt{Financial Children} feature. This consistency suggests that the underlying predictive relationships in Washington's data are similar to the national average, but perhaps with clearer signals or less noise, allowing the model to learn more effectively. The robust performance could also be attributed to a more distinct separation of classes within Washington's dataset, demographics, and financial characteristics that make the predictive task inherently simpler for the model.


In contrast to Washington, Hawaii exhibited a distinct performance signature characterized by a generally low recall value but high precision. \Cref{fig:haw-metrics} illustrates this pattern, where precision often remains high (above 0.6), while recall frequently dips below 0.5, sometimes even falls to 0.2. This indicates that the general model, when applied to the individual state, tends to make many negative predictions that catch the majority of the respondents correctly but in turn misses out on a large amount of positive cases.


Hawaii's SHAP (\Cref{fig:haw-feature}), on the other hand, demonstrates a more skewed impact, with \texttt{Gender/Age Bin (Gender)} and \texttt{Living Arrangements} having a much higher impact compared to the global model, while other features like \texttt{Financial Children}, \texttt{Education}, and \texttt{Health Insurance} have a below-average impact. This is likely due to the client's lower recall score, which causes the model to exhibit less confidence in the features with a more nuanced separation; When a model struggles to identify all positive instances (low recall), it often becomes more reliant on the stronger, most unambiguous signals, and less on subtle contributions from features with smaller or more complex relationships, leading to a more polarized importance distribution.


As seen in \Cref{fig:ny-metrics} and \Cref{fig:ny-feature}, The trends of a typical near-average state like New York generally follow the global model's behavior more closely than the outliers above, with only features like \texttt{Gender/Age Bin (Gender)}, \texttt{Living Arrangements}, and \texttt{Employment} having a noticeable difference. Looking back to the state's metrics, while showing surprising variation, they tend to oscillate around the global model's metrics (\Cref{fig:metrics}). The consistent dominance of \texttt{Gender/Age} features and \texttt{Annual Income} in populated states like New York reinforces the uniformity of these critical demographic and financial indicators in predicting DCA contact across significant populations, highlighting their broad relevance in the predictive task and thus allow the FL model to generalize effectively.

\subsection{Global vs. Local Insights Summary}

A key strength of our FL framework lies in its ability to simultaneously reveal both global (nationwide) patterns and local (state-specific) nuances in financial distress predictors. 

At the global level, the most influential predictors identified by both SHAP and Owen value analysis include demographic features such as gender-age groupings, annual income, and employment status. These variables consistently ranked among the top contributors to model output across most clients, indicating their generalizability as risk signals for consumer financial distress. For example, lower income brackets and unemployment status were associated with higher model-predicted likelihood of DCA contact, aligning with long-standing findings in credit risk literature~\cite{altman2017financial,keasey2019financial,yang2025financial}.

In contrast, state-specific patterns exhibited meaningful deviations from these global trends. For instance, the model performed particularly well in Washington state, where both recall and precision were high, suggesting clearer separability of at-risk respondents. SHAP analysis for Washington mirrored global feature rankings but showed stronger emphasis on employment status and income, indicating that the local data distribution may accentuate these effects. Conversely, Hawaii demonstrated high precision but low recall, and its SHAP profile prioritized different features, such as living arrangements and gender, while underweighting financial dependents and education. These differences suggest that localized economic factors, cultural norms, or even variations in state DCA practices may influence model performance and feature relevance.

Furthermore, Owen values revealed more stable and interpretable patterns for binned categorical features at the state level. For example, grouped features like \texttt{Gender/Age Bin} showed balanced contributions across states with sufficient class diversity, whereas states with low within-group variance (e.g., rural states with fewer younger respondents) showed skewed distributions.

These observed differences open up several plausible policy- and behavior-driven interpretations. For example, Hawaii’s lower recall may stem from its higher cost of living and relatively strong social safety nets, which could result in moderate financial strain without triggering formal DCA, making distress harder to detect through surface-level indicators like employment status or income. In contrast, Washington’s strong performance may reflect clearer reporting patterns or more direct links between income levels and DCA contact. 
Meanwhile, in rural or lower-density states, features like financial dependents or employment status were stronger indicators, likely reflecting household-level vulnerability rather than cost-of-living constraints. These distinctions suggest that interventions, such as financial counseling or hardship assistance, are more effective when tailored to the region's dominant risk factors.

\vspace{-5mm}
\section{Conclusion}


Early identification of at-risk individuals is crucial for enabling proactive interventions by financial institutions and social welfare organizations. 
However, assembling large, centralized datasets for this purpose is often impractical due to privacy concerns, regulatory barriers, and incomplete coverage, making FL a more viable and collaborative alternative for building robust predictive models. Furthermore, FL can be applied on a much larger scale than traditional centralized learning. In addition to addressing problems typically handled by centralized methods, FL can also be implemented across a variety of real-world decentralized systems — for instance, collaborative disease prediction among hospitals or genomic research across laboratories that are unwilling to share raw data directly.
Given the high stakes of financial hardship, explainability is crucial for transparency and trust, especially in policy and aid decisions.

In this work, we demonstrate the viability of FL for predicting consumer financial distress using state-siloed data from the NFCS.
We developed an 8-layer Highway Network architecture tailored to handle highly categorical, imbalanced real-world data, using class weighting to address the significant imbalance (only 18\% of respondents had been contacted by a DCA), and showed through comparison that our results were on-par with the centralized model, while also outperforming localized ones. We also optimized model update communication, reducing costs from 12.954 GB to 4.935 GB over 200 training rounds.
A key contribution of this work is the integration of XAI techniques, such as SHAP and Owen values, to provide actionable insights. Our analysis identified critical predictors of financial distress, including \texttt{Gender/Age} and Annual Income.
This interpretability fosters trust in financial decision-making and supports targeted interventions. 

Our results highlight the value of FL in uncovering regional disparities in financial risk signals without requiring centralized data pooling. This enables policymakers, regulators, and financial institutions to tailor interventions, such as financial education programs, hardship assistance, or regulatory enforcement, based on state-level drivers while still benefiting from the statistical power of a national model. Our approach thus bridges the gap between macro-level analysis and micro-level action, offering a robust framework for equitable, data-driven financial inclusion strategies.

\balance
\bibliographystyle{ACM-Reference-Format}
\bibliography{references}

\end{document}